\documentstyle[aps,epsfig]{revtex}

\begin{document}

\preprint{UTF-417}

\draft

\title{Nucleon Structure Functions and Light-Front Dynamics}

\author{M. Traini$\,^{a,}$\thanks{Talk given at the ECT$\star$/CEBAF joint
Workshop on N$\star$ Physics and Nonperturbative QCD, ECT$\star$, 
Villazzano, (Trento), Italy, May 1998},
Pietro Faccioli$\,^a$,
Vicente Vento$\,^{b,}$}
\address{$^a$ Dipartimento di Fisica, Universit\`a degli Studi di Trento,
and Istituto Nazionale di Fisica Nucleare, G.C. Trento,
I-38050 POVO (Trento), Italy\\
$^b$ Departament de F\'{\i}sica Te\`orica, Universitat 
de Val\`encia,
and Institut de F\'{\i}sica Corpuscular, Centre Mixt 
Universitat de Val\`encia,
Consejo Superior de Investigaciones Cient\'{\i}ficas,
E-46100 Burjassot (Val\`encia) Spain}

\maketitle

\begin{abstract}
We present a quark-parton model to describe polarized and unpolarized 
nucleon structure functions. The twist-two matrix elements for the QCD
evolution analysis of lepton-hadron scattering are calculated within
a light-front covariant quark model. The relativistic effects in the
three-body wave function are discussed for both the polarized and unpolarized
cases. Predictions are given for the polarized gluon distributions as will
be seen in future experiments.
\end{abstract}

\section{Introduction}

The description of deep inelastic lepton-hadron scattering data requires
sophisticated and sometimes ad hoc parametrizations 
showing the complexity of the mechanisms involved in the description 
of the hadron structure. The research reported here analyzes how 
one can visualize these mechanisms in a scheme which unifies the description 
of low and high energy phenomena.
We discuss a radiative approach which makes use of Quark Models (QM) to
calculate the values of the twist two nucleon matrix elements occurring
in the QCD analysis of lepton hadron scattering. QM require a
reinterpretation in order to be used in conjunction with QCD perturbation
theory. Jaffe and Ross \cite{jaffeross80} proposed that the quark
model calculation of matrix elements give their values at a hadronic scale
$\mu_0^2$ and that for all larger $Q^2$ their coefficient functions evolve
according to perturbative QCD. 

Our formalism puts all these ingredients into a predictive scheme 
\cite{noi1,noi2}.
We consider the nucleon to be consistent of valence
quarks and gluons at the hadronic scale $Q^2=\mu_0^2$ and generate 
the partonic content at $Q^2 \gg \mu_0^2$ dynamically via bremsstrahlung 
radiation of gluons and sea from the original system. 
The input distributions are explicitly related to the
electromagnetic response of the constituent quark model which represents the
non-perturbative part of the calculation. 
The investigations developed so far \cite{noi1,noi2,MaTr97,MaTr98,SVT97ss} 
have been based on non-relativistic quark model wave functions. In here
we will demonstrate that the same approach can be 
used to incorporate realtivistic
covariance in a rather transparent way and to this aim
we develop a relativistic quark model making use of the light-front 
hamiltonian dynamics (for reviews cfr. refs. \cite{lf}).

\section{Parton distributions at the hadronic scale $\mu_0^2$}

The parton distributions at the hadronic scale are assumed to be valence quarks 
and gluons, and their twist two component is determined by the quark momentum 
density (cfr. ref.\cite{noi2})

In the light-front quark model
the intrinsic momenta of the constituent quarks ($k_i$) can be obtained from
the corresponding momenta ($p_i$) in a generic reference frame through a
light-front boost 
($k_i = {\cal L}_f^{-1}(P_{\rm tot})\,p_i$, $P_{\rm tot} \equiv \sum_{i=1}^3\,p_i$) 
such that the Wigner rotations reduce to indentities. With the specific choice 
${\cal L}_f^{-1}(P_{\rm tot})\,P_{\rm tot} = (M_0, 0,0,0)$, 
one has $\sum_{i=1}^3 {\bf k}_i = 0$ and $M_0 = \sum_{i=1}^3\,\omega_i =
\sum_{i=1}^3\,\sqrt{{\bf k}_i^2+m_i^2}$. 
The nucleon state is characterized by isospin (and its third
component), parity, light-front (non-interacting) angular momentum operators 
$J$ and projection $J_{\hat n}$, where the unitary vector $\hat n = (0,0,1)$ defines the spin
quantization axis. The nucleon state factorizes into $|N, J, J_n \rangle\,
|\tilde P\rangle$ where $\tilde P$ is the total light-front nucleon momentum
$\tilde P \equiv (P^+, {\bf P}_\perp) = \tilde p_1 + \tilde p_2 + \tilde p_3 $.
$P^+ = P\,^0 + \hat n \cdot {\bf P}$ and the subscript $\perp$ indicates the
perpendicular projection with respect to the $\hat n $ axis. In order to 
achieve the ordinary composition rules, the intrinsic light-front
angular momentum eigenstate $|N, J, J_n \rangle$ must be obatined
from the {\em canonical} angular momentum eigenstate 
$|N, j, j_n \rangle$ by means of a unitary
transformation which is a direct product of generalized
Melosh rotations \cite{Melosh}.
Finally the intrinsic part of the nucleon state, 
$|N, j, j_n \rangle$ is eigenstate of the mass operator 
$
(M_0 + V)\,|N, j, j_n \rangle = M\,|N, j, j_n \rangle
$, 
where the interaction term $V$ must be independent on the total momentum
$P_{\rm tot}$ and invariant under spatial rotations (cfr. refs.\cite{lf}).

In the present work we will discuss results of a confining mass equation of the
following kind
\begin{equation}
\left(M_0 + V\right)\,\psi_{0,0}(\xi) \equiv \left(\sum_{i=1}^3\,\sqrt{{\bf k}_i^2+m_i^2} 
-{\tau \over \xi} + \kappa_l \,\xi\,\right)\,\psi_{0,0}(\xi) = 
M\,\psi_{0,0}(\xi)\,\,,
\label{massop}
\end{equation}
where $\xi = \sqrt{\vec \rho\,^2 + \vec \lambda\,^2}$ is the radius of 
the hypersphere in six dimension and $\vec \rho$ and $\vec \lambda$ are the intrinsic 
Jacobi coordinates $\vec \rho = ({\bf r}_1 - {\bf r}_2)/\sqrt2$, 
$\vec \lambda =({\bf r}_1 + {\bf r}_2 -2\,{\bf r}_3)/\sqrt6$ (solutions for 
non-relativistic reductions of Eq.(\ref{massop}) have been discussed
by Ferraris {\it et al.} \cite{TBM95}).

The intrinsic nucleon state is antisymmetric in the color degree of freedom
and symmetric with respect the orbital, spin and flavor coordinates. In
particular, disregarding the color part, one can write
$
|N, J, J_n = +1/2 \rangle = \psi_{0,0}(\xi)\,{\cal Y}\,^{(0,0)}_{[0,0,0]}(\Omega)\,
\,\left[\chi_{MS} \phi_{MS} + \chi_{MA} \phi_{MA}\right]/\sqrt {2}
$,
where $\psi_{\gamma,\nu}(\xi)$ is the hyperadial wave function solution of
Eq.~(\ref{massop}), ${\cal Y}\,^{(L,M)}_{[\gamma,l_\rho,l_\lambda]}(\Omega)$
the hyperspherical harmonics defined in the hypersphere of unitary radius,
and $\phi$ and $\chi$ the flavor and spin wave function of mixed $SU(2)$
symmetry. Let us note that, in order to preserve relativistic covariance, 
the spin wave functions
have to be formulated by means of the appropriate Melosh transformation of the
i$th$ quark spin wave function:

We have solved the mass equation (\ref{massop}) numerically 
by expanding the hyperradial wave functions $\psi_{\gamma \nu}(\xi)$ on a 
truncated set of hyperharmonic oscillator basis states \cite{TrFa}. 
Making use of the Rayleigh-Ritz variational principle the HO constant
has been determined and convergence has been reached considering a basis as
large as 17 HO components. The parameters of the
interaction, have been determined phenomenologically in order to reproduce 
the basic features of the (non strange) baryonic spectrum up to $\approx 1600$
MeV, namely the position of the Roper resonace and the average value of 
the $1^-$ states\footnote{
			  The well known problem of the energy
location of the Roper resonance is solved, in the present case, 
by the use of $1/\xi$ potential, as discussed in the 
non-relativistic case by Ferraris {\it et al.} \cite{TBM95}.}.
We obtain: $\tau = 3.3$ and $\kappa_l = 1.8$ fm$^{-2}$ \cite{TrFa} to be 
compared with the corresponding non-relativistic fit $\tau = 4.59$ and 
$\kappa_l = 1.61$ fm$^{-2}$ \cite{TBM95}. The constituent quark masses have been
chosen $m_u = m_d = m_q = M_N/3$. 

As a result a huge amount of high momentum components is generated in
solving the mass equation (cfr. Fig.~1.), and they play an important role 
in the evaluation of
transitions and elastic form factors within light-front constituent quark 
models as discussed by Cardarelli {\it et al.} \cite{romalf}
in connection with the solutions of the Isgur-Capstick model Hamiltonian. 

The effects of the high momentum components on the unpolarized parton 
distributions at the
hadronic scale are shown on the right panel of Fig.~1. Their important role 
to reproduce the behaviour of the structure functions for large value
of the Bjorken variable $x$ will be discussed in the next section.
The relevant effects of relativistic covariance are even more evident looking
at the polarized distributions \cite{FaTraVe98}. 
In that channel the introduction of Melosh
transformations results in a substantial suppression of the responses at large 
values of $x$ and in an enhancement of the response for $x \lesssim 0.15$ as
can be seen from Fig.~2.

\section{Numerical results and comparison with the experimental data}
\label{NR&d}

The results we are going to comment are related to two scenarios
according to the assumption on the gluon distribution at the hadronic 
scale $G(x,\mu_0^2)$:

\noindent i) scenario $A$: Quark model or extreme scenario, defined in such a
way that only valence quarks exist at the hadronic scale (i.e.
$G(x,\mu_0^2) = 0$). 
One has \cite{MaTr97}, $\mu_0^2 = 0.094$ GeV$^2$ at NLO
([$\alpha_s(\mu_0^2)/(4\,\pi)]_{\rm NLO} = 0.142$).

\noindent ii) scenario $B$: Partonic scenario, characterized by the existence
of valence quarks and gluons at the hadronic scale.
A natural choice for the unpolarized gluon distribution within the present 
approach, has been discussed in refs.\cite{noi2,MaTr97} and it assumes the 
{\em valence-like} form 
$
G(x,\mu_0^2)= {{\cal N}_g}\,\left[ u_V(x,\mu_0^2) +
d_V(x,\mu_0^2)\right]/3\,\,.
$
As a consequence $\int G(x,\mu_0^2)\,dx = 2$ and only $60\%$ of the total 
momentum is carried by the valence quarks at the scale $\mu_0^2$.

If the gluons were fully polarized one would have 
$|\Delta G(x,\mu_0^2)| = G(x,\mu_0^2)$, which reduces to
$
\Delta G(x,\mu_0^2) = f \,G(x,\mu_0^2)
$
introducing the fraction $f$ of polarized gluons, to be considered 
with the appropriate sign. As an example we discuss results followig 
a suggestion due to Jaffe \cite{jaffe96}: $f \approx -0.35$
($\int dx\,\Delta G(x,\mu_0^2) \approx -0.7$). In this case 
one obtains a consistent lower bound to the $x$-dependence of 
$\Delta G(x,\mu_0^2)$
\footnote{In fact in ref.\cite{jaffe96} it has
been shown that $\int \Delta G(x,\mu_0^2)\,dx < 0$. Such inequality 
does not imply $\Delta G(x,\mu_0^2) < 0$ in the whole $x$-range. 
We are therefore investigating a lower bound to $\Delta G(x,\mu_0^2)$.}. 
In this case $\mu_0^2 = 0.220$ GeV$^2$ at NLO
([$\alpha_s(\mu_0^2)/(4\,\pi)]_{\rm NLO} = 0.053$).

In Fig.~3 the results for the proton structure function $g_1^p(x,Q^2)$ are
shown and compared with the experimental data within scenario $A$. 
The non relativistic
approximation appears to reproduce rather poorly the experimental observations,
a result already discussed in ref.\cite{MaTr97} for other non-relativistic
quark model wave functions. The introduction of 
relativistic covariance in the quark wave function, mainly due to spin
dynamics induced by the Melosh rotations \cite{FaTraVe98}, leads to 
a suppression of the structure in the small-$x$ region ($x \lesssim 0.5$), 
Such a large effect brings the theoretical predictions quite close to
experimental data in the region $0.01 \leq x \lesssim 0.4$, even under the
simple assumption of a pure valence component at the hadronic scale 
(scenario $A$). 
We stress that the calculation is parameter-free and the only adjustable 
parameters ($\tau$ and $\kappa_l$ in Eq.~(\ref{massop})) have been fixed 
to reproduce the low-lying nucleon spectrum as already discussed.

Let us comment also on the comparison of LO versus NLO calculations. 
The differences shown in Fig.~3 indicate the relevance of higher order 
corrections in our parton model approach. The initial scale $\mu_0^2$ is 
rather low, and NLO corrections have to be included.

In order to introduce gluons we evolve the unpolarized distributions predicted 
by the scenario $A$, up to the scale of scenario $B$ where 60\% only of 
the total momentum is carried by valence  partons.
At that scale the fraction of polarized gluons is chosen to be negative, 
according to the Jaffe result \cite{jaffe96} (scenario $B$).
Looking at the Figs.~3,4 one can conclude that the low-$x$ data on 
$g_1^p$ do not constrain the gluon strongly. 
If the fraction of polarized gluons varies from 35\% to 100\% the 
quality of the agreement is  deteriorated in the region 
$0.01 \leq x \lesssim 0.4$ only slightly. For larger values of $x$ the valence 
contribution plays a major role and the behaviour of the structure functions 
will depend largely on the potential model.

The comparison of the predicted neutron structure function with
the data (Figs.~3,4) differs quite substantially according to the amount of 
polarized gluons at the hadronic scale. 

Within scenario $A$ the values of $x\,g_1^n(x,Q^2)$ remain quite small 
according to the fact that the mass operator
(\ref{massop}) is $SU(6)$ symmetric, while the introduction of {\em negative}
gluon polarization, as suggested by Jaffe, brings the predictions of the present
relativistic quark model quite close to the experimental observations at least
in the $x \gtrsim 0.1$ region. Larger {\em negative} fraction of gluon 
polarization is favored by the data in agreement with the large effect required
to split the nucleon and $\Delta$ mass at the hadronic mass scales.

In Fig.~5 we show results for the unpolarized $F_2^{\rm p,n}$
structure functions. The relativistic approach improves their description 
in the whole $x$-range, in particular in the region $x \gtrsim 0.4$ where
high momentum components play a relevant role and the valence distribution
dominate the response. In the complementary range ($x \lesssim 0.4$) 
the distributions could be improved by considering the non-perturbative 
$q \bar q$ contributions (cfr.ref.\cite{MaTr98}) neglected at the present 
stage of development of our relativistic scheme.

Finally  in Fig.~6 we show the gluon ditributions both for polarized and
unpolarized scattering. $\Delta G$ will be measured in dedicated experiments 
at CERN and BNL as discussed also during the workshop and our predictions
show the sensitivity of that observable on the gluon polarization at the scale
of the constituent quark model

\protect
\begin{figure}
\begin{center}
\mbox{\epsfig{file=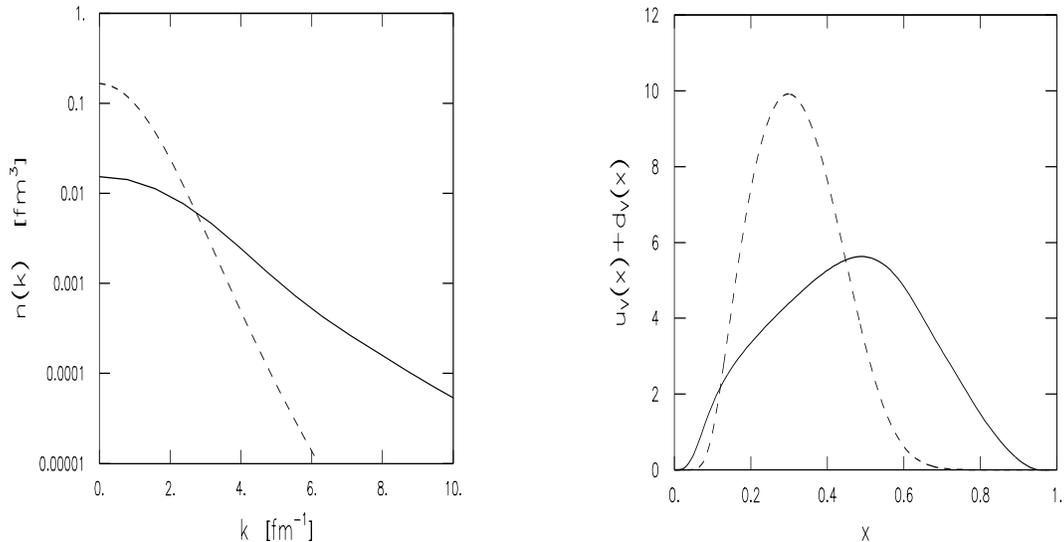,
width=0.4\linewidth,height=0.60\textheight, angle=-90}}
\end{center}
\caption{Left panel: $n({\bf k}^2) = \sum_q [n_q^{\uparrow}({\bf k}^2) +
n_q^{\downarrow}({\bf k}^2)]$ (the valence quark momentum distributions)  
as function of $|{\bf k}|$.
Relativistic results: full curve, non-relativistic approximation: dashed curve.
On the right panel the corresponding total valence distributions 
$u_{\rm V}(x,\mu_0^2) + d_{\rm V}(x,\mu_0^2)$, at the hadronic scale.}
\end{figure}

\protect
\begin{figure}
\begin{center}
\mbox{\epsfig{file=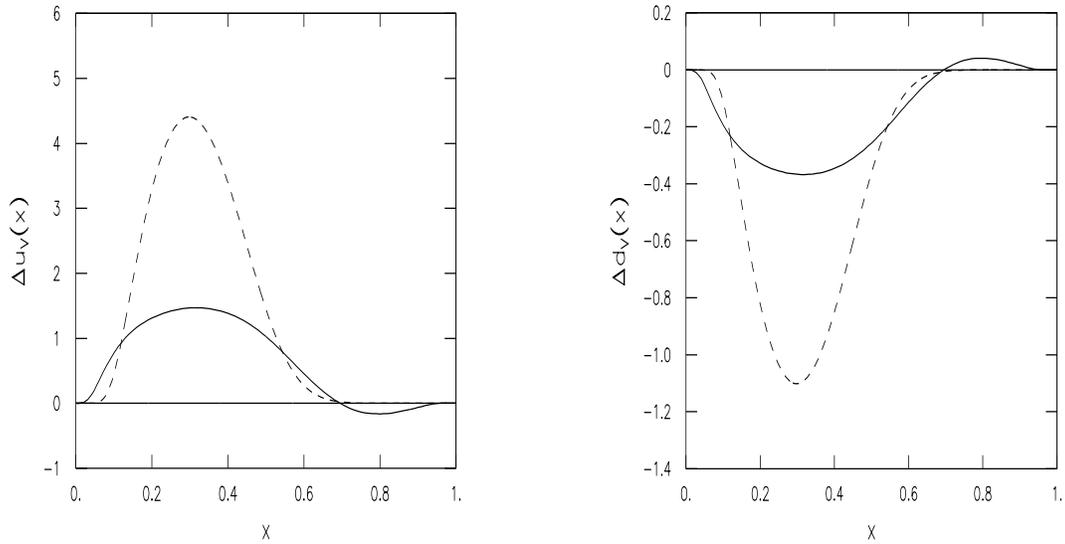,
width=0.4\linewidth,height=0.60\textheight, angle=-90}}
\end{center}
\caption{Left panel: the polarized distribution $\Delta u_{\rm V}(x,\mu_0^2)$ 
as function of $x$ within the relativistic (full curve) and 
non relativistic (dot-dashed curve) schemes.
On the right panel the distribution $\Delta d_{\rm V}(x,\mu_0^2)$
same notations as in Fig.~1.}
\end{figure}

\protect
\begin{figure}
\begin{center}
\mbox{\epsfig{file=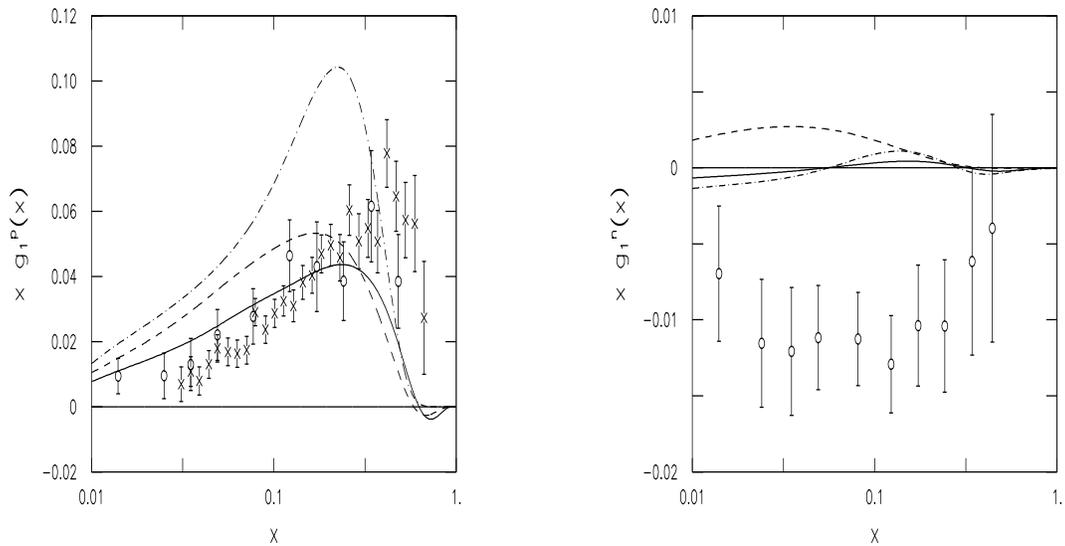,
width=0.4\linewidth,height=0.60\textheight, angle=-90}}
\end{center}
\caption{The proton (left) and neutron (right) polarized structure functions at
$Q^2=3$ GeV$^2$, within scenario $A$. The full curves represent the
relativistic results abtained by means of a complete NLO evolution; the dashed
curves show the corresponding LO predictions. Dot - dashed curve: the (NLO) non
relativitic calculation. Data are from the SMC and E143 experiments for the
proton [14], and E154 for the neutron [15].}
\end{figure}

\protect
\begin{figure}
\begin{center}
\mbox{\epsfig{file=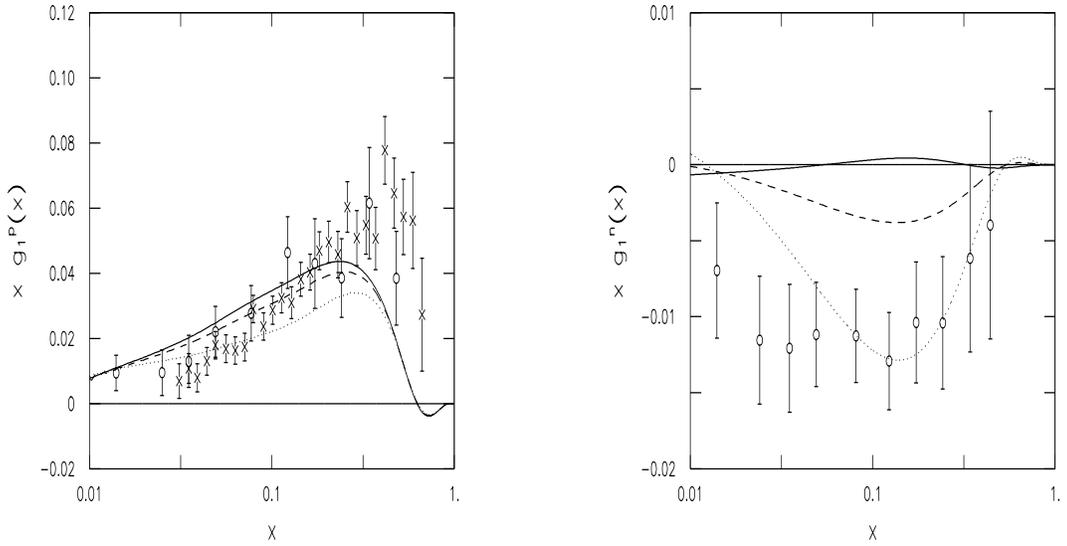,
width=0.4\linewidth,height=0.60\textheight, angle=-90}}
\end{center}
\caption{NLO $x\,g_1^{\rm p,n}(x,Q^2=3$ GeV$^2$) within different scenarios
(see text). Scenario $A$ ($f = 0$): full curves 
Scenario $B$: for different values of the fraction of polarized gluons at 
the hadronic scale: $f=-0.35$ (dashed curves), $f=1$ (dotted curves).
Data as in Fig.~3.}
\end{figure}

\protect
\begin{figure}
\begin{center}
\mbox{\epsfig{file=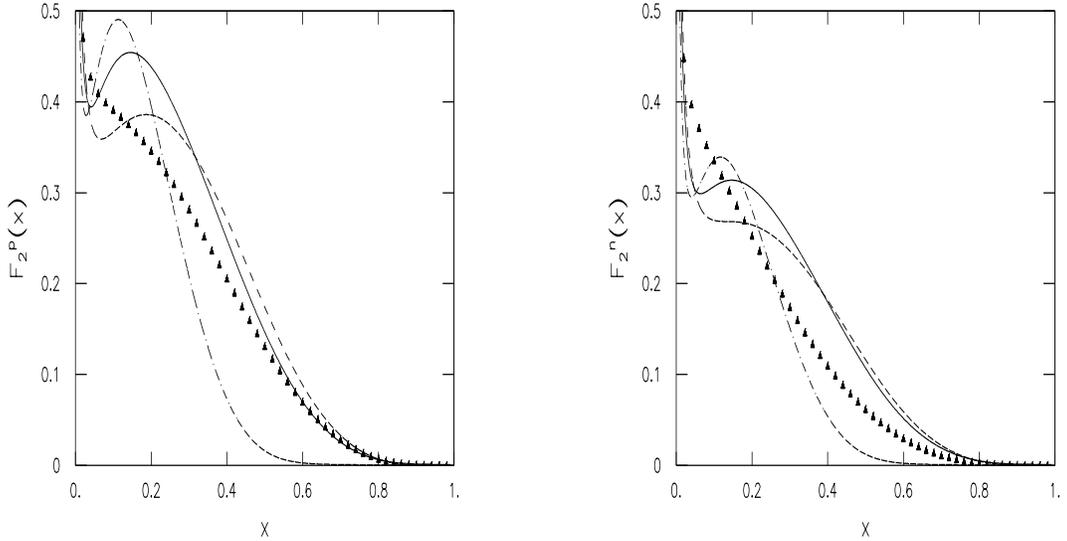,
width=0.4\linewidth,height=0.60\textheight, angle=-90}}
\end{center}
\caption{$F_2^{\rm p,n}(x,Q^2=3$ GeV$^2$) within scenario $A$. 
Notations as in Fig.~3. Data fit (triangles) from ref.[16].}
\end{figure}

\protect
\begin{figure}
\begin{center}
\mbox{\epsfig{file=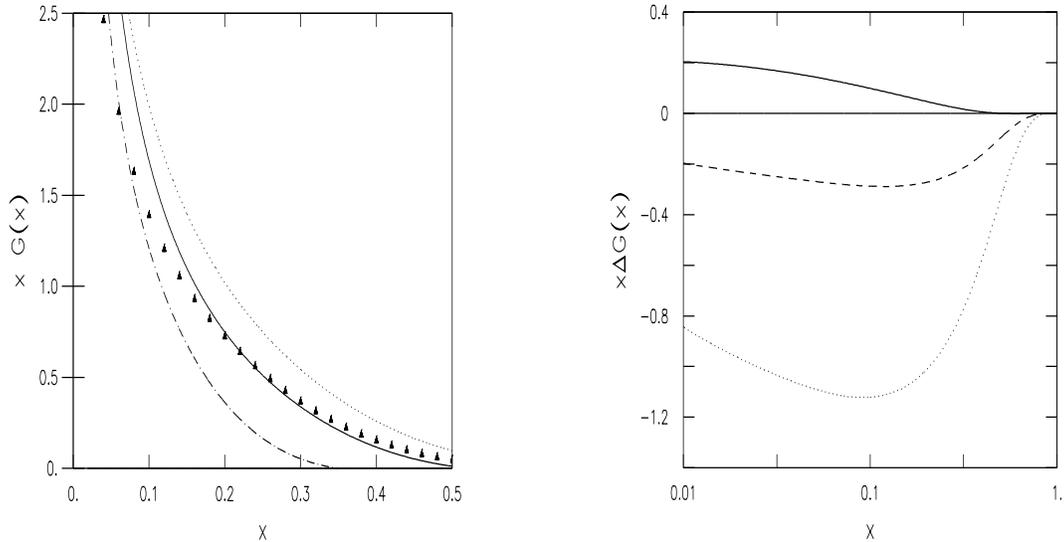,
width=0.4\linewidth,height=0.60\textheight, angle=-90}}
\end{center}
\caption{NLO unpolarized (left) and polarized (right) gluon distributions at
$Q^2 = 3$ GeV$^2$.
Left panel: scenario $A$: relativistic results (full curve), 
non relativistic approximation (dot-dashed curve). Scenario $B$: dotted curve
(DIS factorization scheme). Right panel: scenario $A$: full curve; scenario
$B$ ($\overline {\rm MS}$ factorization scheme): 35\% polarization fraction 
(dashed curve), 100\% polarization fraction (dotted curve).}
\end{figure}


\end{document}